\documentstyle[12pt,twoside,fleqn,espcrc1,epsfig]{article}
\title{ What have we learned and want to learn\\ from heavy ion collisions at CERN SPS?}
 \author{ E.V.~Shuryak\address{State University of New York, 
       Stony Brook, NY 11794, USA}}
\begin{document} 
\maketitle
\begin{abstract}
  The talk is a 
mini-review of the current status of the field, with
emphasis on SPS heavy ion program, now and
 beyond 2000 (as  asked by the organizers).
The main question is, of course, 
whether we can convince ourselves 
and  the community at large
that the QGP is in fact produced at SPS.
  We came a long way toward the  {\em positive} answer, and are
definitely on strongly rising part of the  learning curve. Still,
 in few key directions we lack important pieces of evidences. 
\end{abstract}
\section{The QCD Phase Diagram}
\subsection{Theoretical progress }
\begin{figure}[h]
\begin{center}
\vskip -.5in
 \begin{minipage}[c]{3.in}
 \centering
 \includegraphics[width=2.in, angle=270]{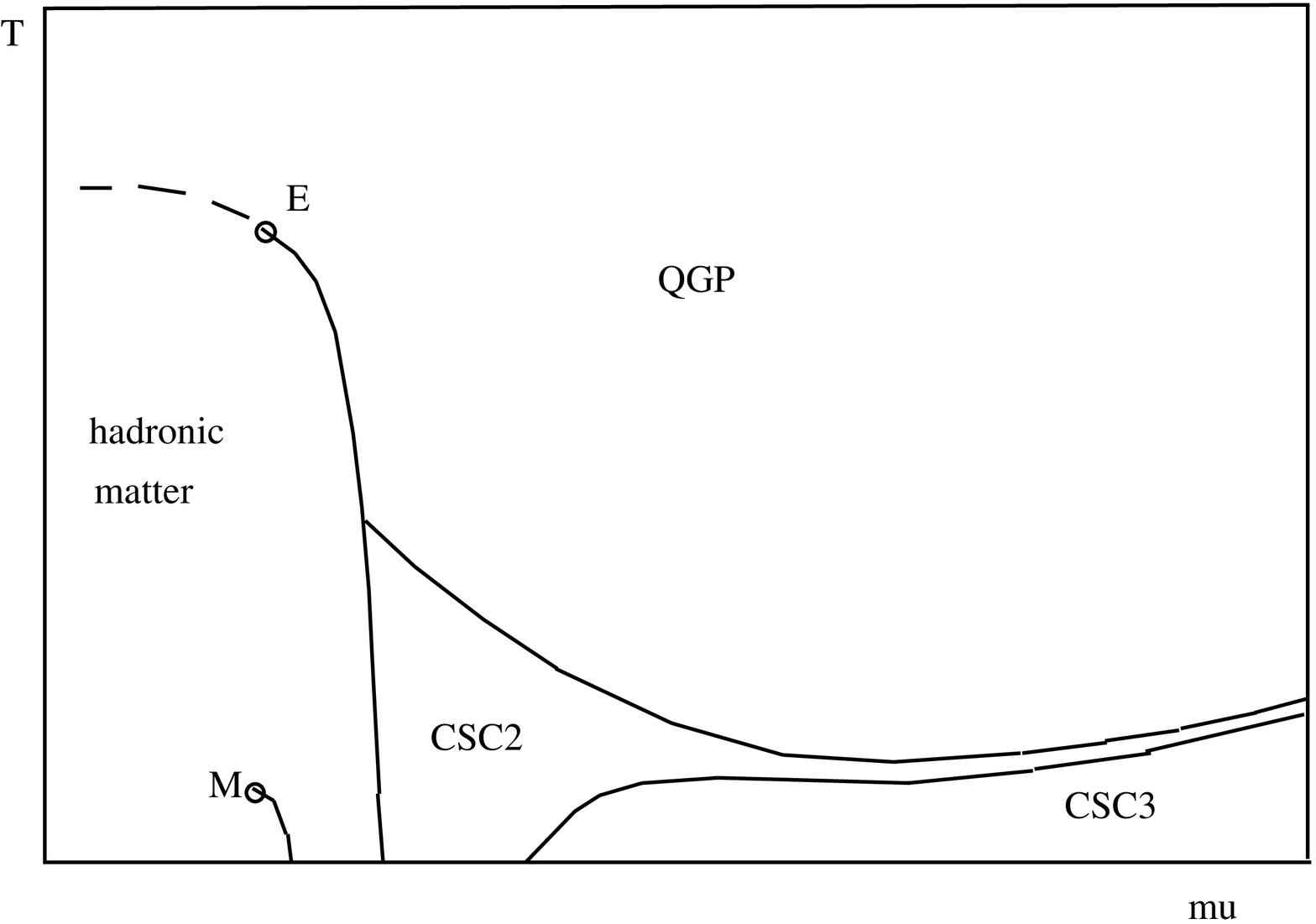}
 \end{minipage}
   \begin{minipage}[c]{3.in}
\vskip -.5cm
   \centering
\includegraphics[width=2.5in, angle=270]{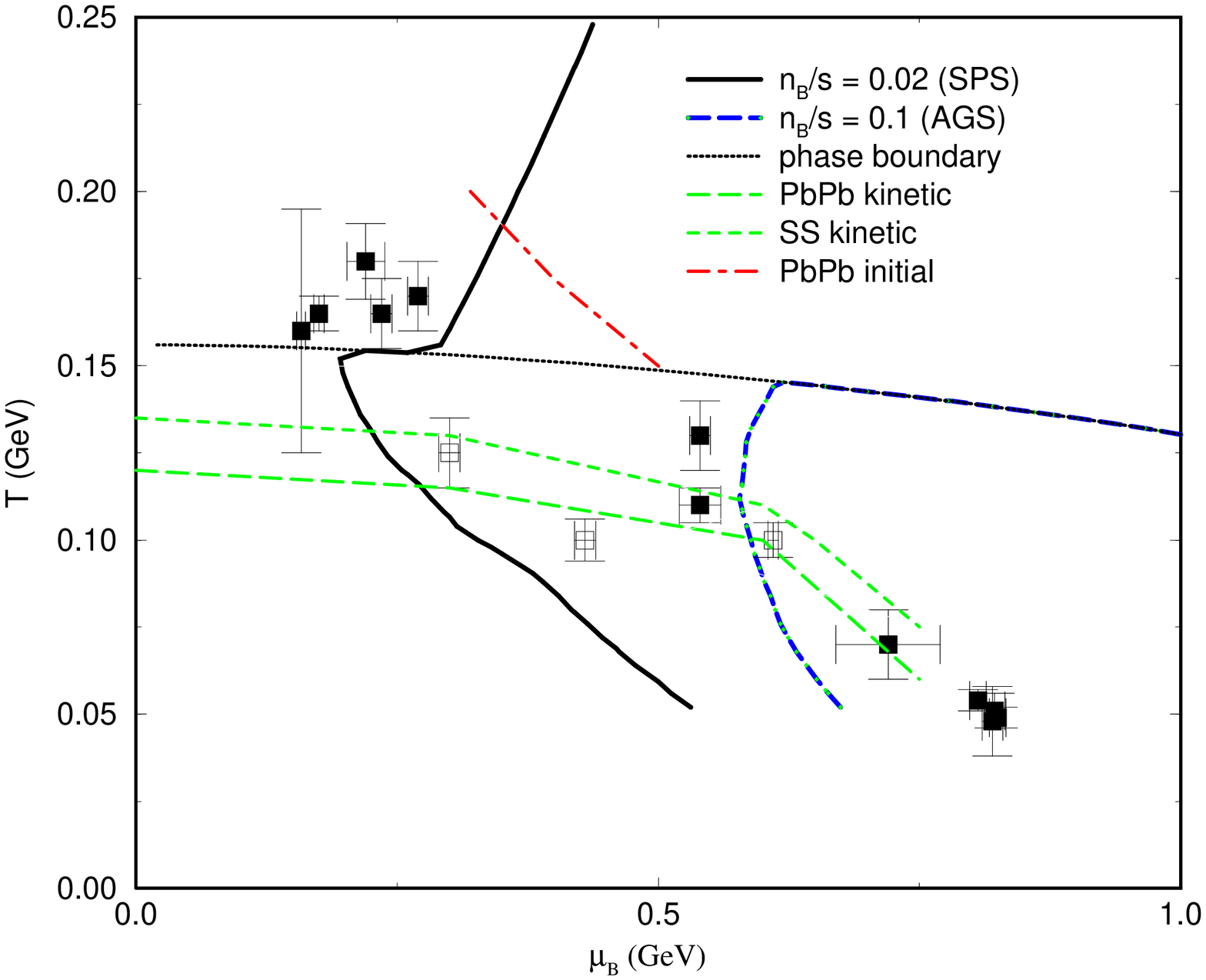}
 \end{minipage}
\vskip -0.5in
\end{center} 
\caption{
\label{fig_phases_th}
(a) Schematic phase diagram of QCD, in temperature T- baryon chemical
potential $\mu$ plane. E and M show critical endpoints of first
order transitions: M (from multi-fragmentation) is that for liquid-gas
transition in
nuclear matter. The color superconducting phases, CSC2 and   CSC3 are
explained in the text.
  (b) The experimentally accessible part of the phase diagram. 
 Closed (open) points correspond to chemical
  (thermal) freeze-out (from analysis compiled in \protect\cite{CR})
The adiabatic
paths correspond to entropy per baryon ratio indicated,
from \protect\cite{HS} . Lines of thermal freeze-out for central  SS and PbPb
collisions,
as well as the
initial line for  PbPb are also indicated.
 }
\vskip -0.2in
\end{figure}

  The  QCD phase diagram version circa 1999  \cite{RSSV2}, shown in Fig
\ref{fig_phases_th}(a), looks rather different from what was shown at
previous Quark Matter conferences.
Most significant progress is seen in the 
large-density low-T region, where two new 
 Color Super-conducting phases have appeared.  Unfortunately
  heavy ion collisions do not cross this part
of the phase diagrams:  it belongs to neutron star
physics.

 The subject is
  covered
in talks \cite{RS}, so I only make
few  comments here.
The 2-flavor-like color superconductor
CSC2 phase was known before \cite{super}, but realization that
it should  be induced by instantons \cite{super_inst} has increased 
the gaps (and $T_c$) from a few MeV
scale to $\sim$100 MeV (50 MeV). It is hardly surprising, since the $same$ interaction in $\bar
q q$ channel is responsible for chiral symmetry breaking, with the gap
(a constituent quark mass) 350-400 MeV.
The symmetries of the CSC2 phase are similar to
the 
electroweak part of the Standard Model, with scalar isoscalar $ud$ diquark  as
Higgs. The colored condensate breaks the color group,
making 5 out of 8 gluons massive.
The 3-flavor-like phase, CSC3, is brand new: it was proposed in \cite{ARW2} based on
one gluon exchange interaction, but in fact it is favored by
instantons
as well \cite{RSSV2}. Its unusual features include {\it color-flavor
  locking}
and {\it coexistence} of both types of condensates, $ <qq>$ and $< \bar q q>$.
It combines features of the Higgs phase (8 massive
gluons) and of the usual hadronic phase (8 massless ``pions'').
Surprisingly, at very large densities Cooper pairs are bound magnetically
\cite{Son},  leading to growing gaps  at extremely large $\mu$.

 Another important new element is  the (remnant
  of) the QCD tricritical point \cite{SRS}.  
Although we do not know where it is\footnote{Its position is very sensitive to
precise
value of the strange quark mass $m_s$}, we hope we know how to find it, see
\cite{SRS}. All ideas proposed rotate around the fact that the
 order parameter, the famous sigma meson, is
at this point truly massless, and creates a kind of a ``critical opalecence''. 

\subsection{ The Phase diagram as our Map}

  The major goal of the heavy ion program in general is to reproduce as
  much as possible properties of dense/hot matter in bulk, and study
 qualitatively new
 phenomena like phase transitions. 
 Now, with  accurate statistical description of the particle
 composition,
 small and Gaussian event-by-event
(EBE) fluctuations, and
detailed  data on collective flow,
people are using macroscopic language more than before, and so
it is entirely appropriate to start with a discussion of {\it where we
  are
on the phase diagram}.

  Looking at the fireball by a detector is like
 looking at the Sun: one can only see the photo-sphere,
with
$T\sim 6000^o$, not the hotter interior. 
Because
 we can see many hadronic species,
we can also trace down conditions at which the composition
is formed.
The result 
 can be summarized as two separate freeze-out
points, $thermal$   or kinetic one, and $chemical$: a recent sample of those
coming from SPS, AGS and SIS data  is shown 
in Fig.\ref{fig_phases_th}(b).  The chemical point are about the same
for all A, but the
thermal ones are ``cooler'' for larger A.
For a discussion of freeze-out systematics see \cite{CR}.
  Only at SPS we find that the
two  freeze-outs are well separated, and so
  we indeed found  a new phase of matter at SPS, never
  seen before: namely a 
   {\it  chemically frozen  thermally equilibrated ``resonance'' gas}.  

  The zigzag shaped
paths on the phase diagram are {\it adiabatic curves} for a particular
EOS \cite{HS}(b) which
I have discussed at QM97 \cite{HS}(c):  one can see that
measured freeze-out points at SPS and AGS happen to
be close to the predicted paths.
  Do
 chemical and thermal  points indeed follow the
same adiabates of ``resonance gas''? The points in Fig.\ref{fig_phases_th}(b) are not
accurate enough to say that, but it can be studied in models.
In \cite{UrQMD} the URQMD cascade was studied\footnote{This model of
  course has no zigzag and QGP, only the hadronic part of the path is
  the same. }, and although
the non-equilibrium effects in some observables can be large,
the exact 
and equilibrium expressions for entropy differ
by only about 6\%. 
 Also the kinetic models agree that most of the entropy is produced 
 early, and then it is only slightly 
affected by re-scattering.  
In summary: the adiabatic paths seem to be the paths to follow!

\section{Hadronic stage}
\subsection{The radial  flow}

 The main news from the last few years is that
at AGS/SPS energy domain the heavy ion collisions really produce a {\it
  Little Bang rather than a  fizzle},
with
strong  explosive transverse flow converting 
a significant part of thermal energy into that of 
collective motion.

 Looking two decades ago at the pp ISR data I have found  \cite{SZ}
  no trace of transverse radial flow: the $m_t$ slopes
for various secondaries were identical.
Only in mid-90s a significant difference in  slopes was
observed, first for  light- and then heavy-ion collisions.   
At all previous QM conferences the origin of these slopes
  was  debated. 
Are   differences in AA
and pp   due to
 ``initial state'' parton re-scattering or  
the final state re-scattering of hadrons? 
Now the debate is mostly over since
the amount of evidences which proves the latter is overwhelming. Let me
mention  two of them.
(i) Initial state scattering may 
  broaden  the nucleon $m_t$ spectra, but it  then predicts
{\it the same} slope for deuterons, contrary to observations. Only a
correlation between the two nucleons can explain data, 
and the calculated flow   reproduces it well\cite{deuteron}.
(ii)  slopes for $\pi,K,N,d$  depend linearly on particle mass
 (common flow velocity $v_t$), except for strange 
baryons.
The largest deviation is found for $\Omega^-$:
 this is nicely explained
\cite{Sorge_omega} by its early freeze-out due to smaller cross
sections\footnote{Why  does the $\phi$ not have a small slope as well? See discussion below. }.  

 How large is the transverse flow velocity $v_t$?
The
 fits to ``hydro-motivated'' formulae   have produced
widely varying  
values, which also were 
 strongly A- and rapidity dependent. The explanation
   was  worked out in the hydro-kinetic framework
  \cite{HS}, which I also described at QM97. 
The $motto$ of this work,
``the larger the system, the further it cools'',
  explains  both strong A and y dependence of flow. Very large
  $v_t\approx .6$ and very low $T_{thermal}=110-120 MeV$ in PbPb were
  predicted\footnote{As only very elementary kinetics of low energy pion
    and nucleon re-scattering is actually involved,  any event
    generator
like RQMD  also ``knew'' about it. Just  it was not put
in the proper words.}.
Later analysis based heavily on NA44,NA49 {\bf HBT radii} $and$  spectra have confirmed such  
   selection \cite{HW}.

    The value of $v_t$ is important because it {\it tells
us  about the EOS of
  hadronic matter}. The observed
values are mostly  generated  by 
``resonance gas'', which at SPS has simple EOS  \cite{resonancegas}  
$p\approx .2 \epsilon$, $ p,\epsilon\sim T^6$.  $v_t$ is large because
``resonance gas'' at $T=120-160 MeV$ has no Hagedorn
``softening''\footnote{The Hagedorn  phenomenon caused by the
  exponential rize of spectral density
 was in fact found to exist in QCD without quarks. It  
accurately explains lattice  data on the value
of the deconfinement transition for $N_c=2,3,4$. So, the Hagedorn
softening would have happened in QCD at $T_{Hagedorn} \approx 260
MeV$, if not for the light-quark dominated transition at
lower T.}. The mixed phase is however softer, with   
   the minimum of $p/\epsilon$ (the {\it the softest point})
corresponding to 
 all matter  converted to QGP. It was predicted \cite{HS}(a) that the 
collisions which start from this condition should
 last longer.
 We are waiting for the next SPS run, at 40 GeV, to
see if this prediction is indeed  confirmed.

\subsection{ Elliptic flow}
In high energy collisions
the shape of the ``initial almond'' for non-central
collisions leads to enhanced ``in-plane'' flow, in the direction of
the impact parameter \cite{Ollitrault}. 
It is very
 important because (as  pointed out 
in \cite{Sorge}(a)) it is developed $earlier$
than the radial one, and thus it may shed  light on whether we do or do not
have QGP at such time.
Now it is  measured by
the asymmetry of the particle $number$, or 
$v_i$ harmonics defined as
$	{dN \over d\phi} = {v_0	\over 2\pi } + {v_2 \over \pi } \cos( 2\phi ) + 
				   {v_4 \over \pi }  \cos( 4\phi ) +
                                   \cdots $
rather than asymmetry of the momentum distribution. 
Furthermore, $v_i$ can be additionally
normalized to the spatial asymmetry of the initial state (the
``almond'')
at the same $b$, in order to cancel out this kinematic factor and to see
the response to asymmetry.

\begin{figure}[h]
\begin{center}
\leavevmode
\epsfxsize=5cm
\vskip -0.7in
 \begin{minipage}[c]{3.in}
 \centering 
\includegraphics[width=2.95in, angle=270]{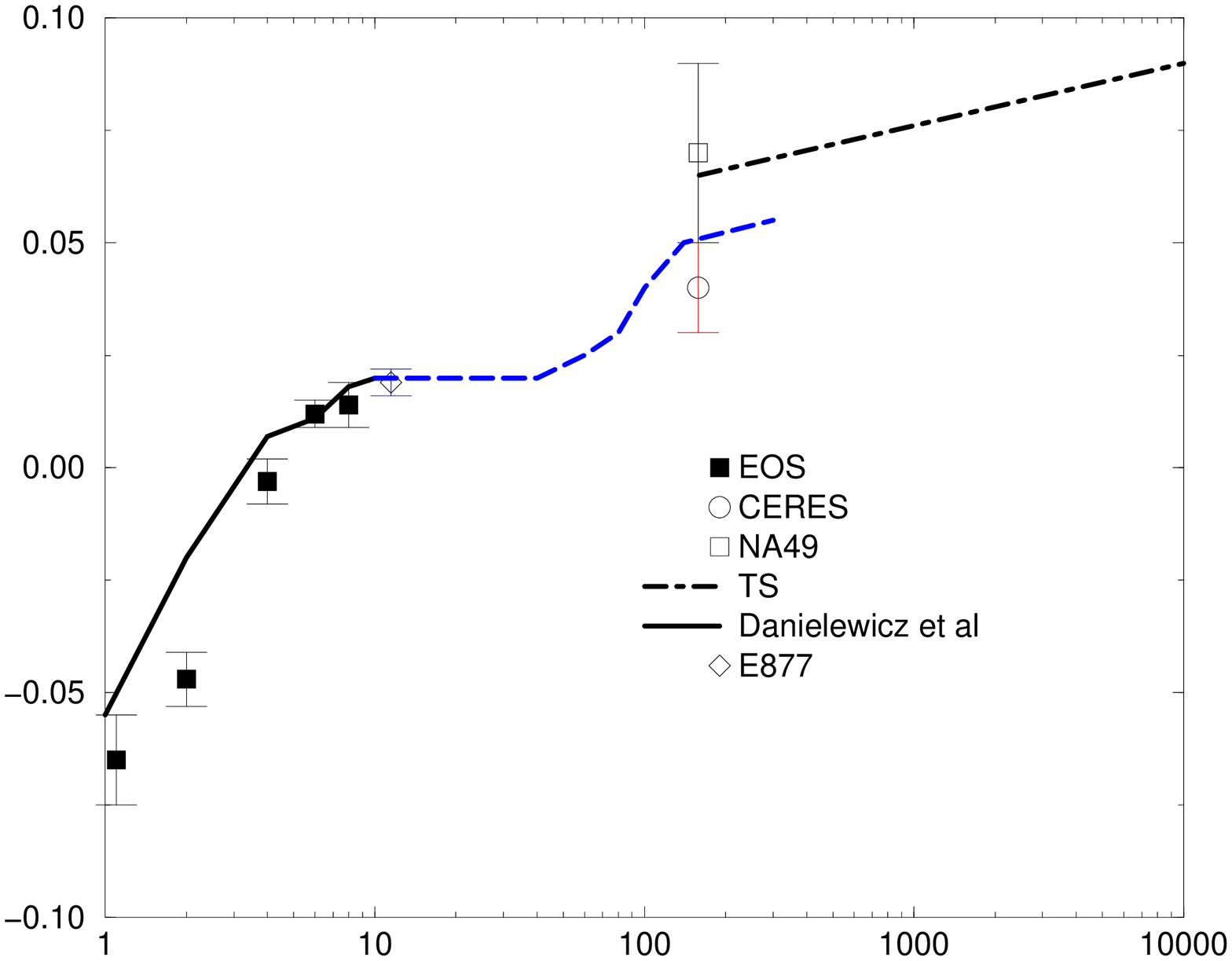}
 \end{minipage}
   \begin{minipage}[c]{3.in}\vskip 0.3cm
   \centering
\includegraphics[width=2.75in]{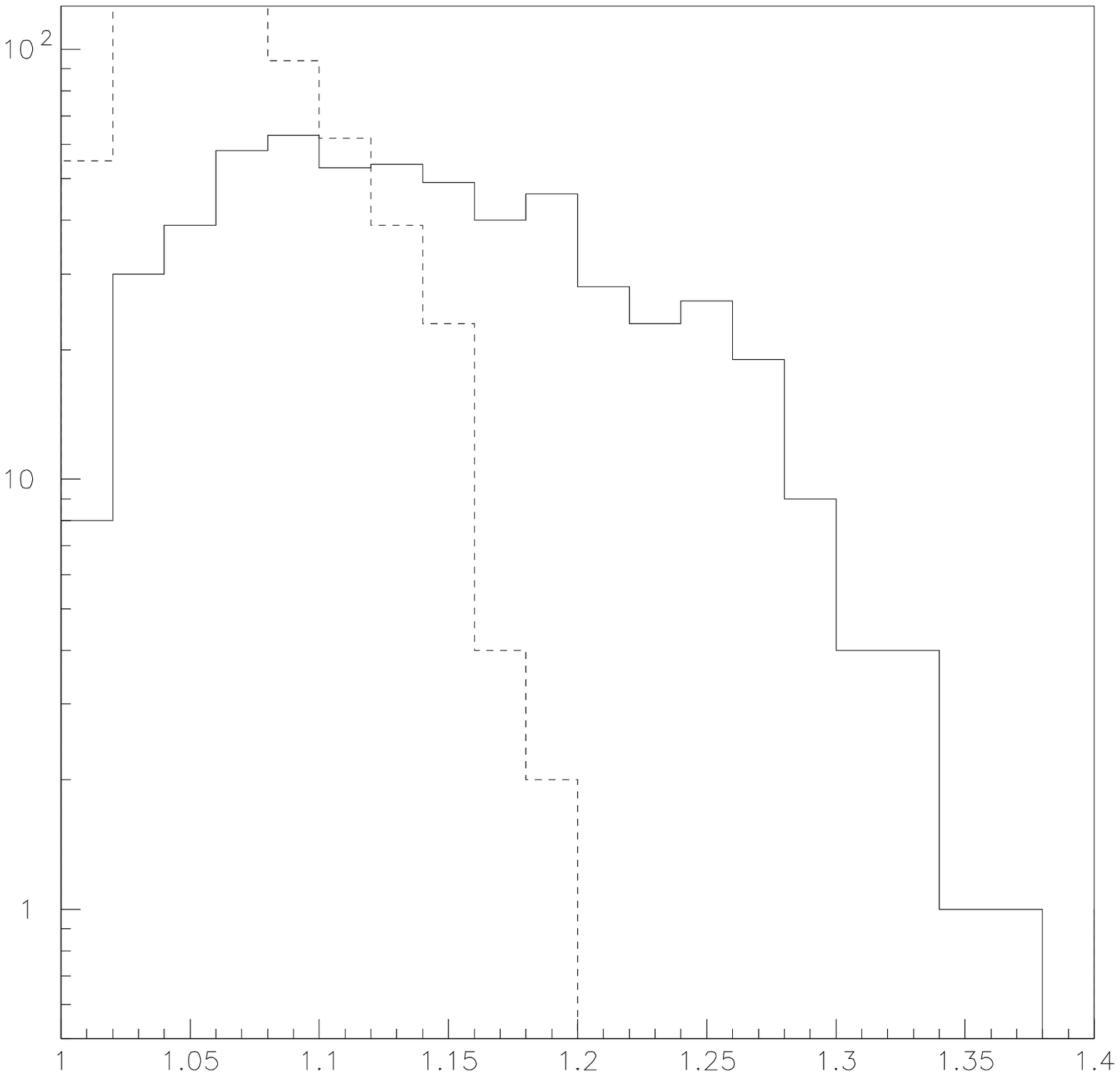} 
\end{minipage}
\vskip -0.5in
\end{center}
\caption{
\label{fig_elliptic}
 Nucleon ellipticity $v_2$ at mid-rapidity and mid-central
collisions, versus collision energy
  [Gev/N]. The experimental sources are (compiled in \protect\cite{Stachel}) and
indicated on figure, TS  is hydro calculation \protect\cite{TS} ,
 the dashed line
   is just a guess (see text).
(b) Distribution of the ratio of long-to-short r.m.s. radii of the
participant nuclei in transverse direction, for large number of participants
$N_p>.9 N_{max}$.
Solid (dashed) histogram is for  UU
(PbPb) collisions.
}\vskip -0.2in
\end{figure}

  Let us first look at the energy dependence:
a compilation of  measured nucleon $v_2$  is shown in
Fig.\ref{fig_elliptic}(a). 
  First news is  indication for {\bf softening of EOS} at 
$E>6 AGeV$ observed by EOS detector at AGS, 
exactly where the initial conditions are expected to hit for the first
time the  critical line (see \cite{Danielewicz}).
 At the same energy
$K/\pi$ ratio and other strangeness enhancement signals change
rapidly.
 Both indicate  that  the mixed phase is
actually  reached {\it already at such low energies}.

 The dash-dotted  curve for  energies above 
 the SPS is from 
hydro calculations  \cite{TS}.  In  agreement with \cite{Ollitrault},
 large and
     near constant ellipticity in this region is found, driven by {\it the
QGP push} at early times.
Strong decrease of this effect is expected in the SPS domain, as the ``QGP
push"  disappears. The dashed curve is my speculation of what the
excitation function of $v_2$ may be in the SPS domain. The existence of 
a $plateau$ is not obvious, but an {\bf inflection point} 
 seems inevitable\footnote{ This prediction looks similar to
 that made  by van Hove \cite{vanHove}
 for radial flow.
His argument was however too naive, because it ignored  time
development: soft EOS leads to longer expansion time.
Therefore
 even larger velocity can be obtained by the end.
But ellipticity is a small effect which does $not$ determine the
expansion time: so  in this case the
argument should be valid.}.

 Let us now turn to  SPS data, which depict $centrality$ dependence of
 $v_2$. This issue is more complicated, because by increasing $b$ we
 make the
``almond'' more elliptic, but also much smaller and thinner: eventually
finite size corrections reduce the pressure build-up\footnote{And this is
  why the guessed dashed curve in Fig.\ref{fig_elliptic}(a)
is below the hydro prediction. }. One needs cascade codes to study this
effect: see \cite{Sorge}. 
  Last year  RQMD has been radically changed,
  including the possibility to vary the EOS and to include the ``QGP push".
 New version \cite{Sorge}(b) predicts a relative
 growth of  (properly normalized to
spatial asymmetry) $v_2$ at small $b$, to values close to those
 predicted by hydro.
 Preliminary 
NA49 data presented at this meeting \cite{NA49} 
indeed observe such enhancement at small $b$. 
 
 Much more work is clearly needed to understand this complex interplay of
EOS and finite size effects. U collisions discussed below in principle
provide the means to decouple finite size and deformation issues.
The next run (40 AGeV) data are also of great
 importance here: no ``QGP push'' is expected there.
 If a clear difference in $v_2(b)$ 
between 40 and 158 GeV is observed, it would really be
 the first sign of {\it the QGP
 push} at SPS. 

 Whether it is seen   at SPS or not, it certainly
is expected to lead to quite
 dramatic phenomena at RHIC.
  A very non-trivial and
highly EOS-sensitive expansion pattern is predicted
in \cite{TS}: 
 The so called
  ``nutcracker'' scenario (for non-central
 collisions) includes
formation of
 two ``shells'',
 which are physically separated from each other
by freeze-out, with only a
 small
``nut'' at the center.  This  behavior predicts  higher
moments $v_4,v_6$ of specific kind
growing with energy, and spectacular HBT radii. 
We are looking forward to first RHIC data to see if this is true.

\subsection{The event-by-event fluctuations:  all events are (about) equal! } 
   Search for ``unusual'' events (e.g. 
 disoriented chiral condensate) had
 attracted significant attention in the past, 
but they were not found.  NA49 has shown that 
fluctuations in such observables as $<p_t>$ are rather small and 
 Gaussian,
without unusual tails. Their widths can be measured rather accurately.
What can we learn from them?

  It is tempting to
 apply  thermodynamical approach based on entropy here\footnote{
Such
type of analysis of multi-hadron production reactions goes back to 70's, see
e.g. my own work \cite{ES_72} where
probabilities of  exclusive channels for low energy
 $\bar p p$ annihilation were calculated from ideal gas 
entropy.}. Furthermore,
as the temperature fluctuation is related to specific heat
$\Delta T^2/T^2=1/C_v$,
 it was proposed in \cite{Cv} to use it to measure
 $C_v$ of a hadronic matter at freeze-out.
For example, at the critical point mentioned above, $C_v$ diverges
and $\Delta T$ must vanish. Can this dramatic prediction
 be somehow observed?
Unfortunately, as explained in \cite{SRS}, this argument is too naive.
 Pions can indeed be used as
a ``thermometer", in contact with a
 fluctuating medium (the sigmas), but fluctuations of its T 
 measures mostly the {\it thermometer's own} $C_v$. The
critical fluctuations can be seen, but only as a correction at the
10\% level or so.

 The fluctuations  for intensive
and extensive observables are generally of different nature.
 An example of the former case
is $<p_t>$: the deviation of the measured width from pure statistics
(mixed events) is in this case
surprisingly small\footnote{Even Bose enhancement
  which definitely is there in HBT measurements is 
 canceled out.}.
Fluctuations of particle composition \cite{Mrow} are also intensive,
and they are sensitive to resonances.
New point: if one would be able to get any hold on
multiple production of multi-strange objects, one can
tell if such exotic objects as ``color ropes'' do or do not exist.

 An example of an extensive variable
 is total multiplicity: even restricted to central 5\% NA49 finds
a Gaussian width  $twice$  that
for random (Poisson) emission. As
 shown in \cite{SRS}, only about half of the effect comes from
  resonances at freeze-out. The rest must come from
fluctuations in initial conditions: see 
 discussion in \cite{BH}.
 I
think the most interesting part of the development is 
attention to mixture of the two, such as $<\Delta p_t \Delta
N>$. As explained in \cite{SRS}, those do not appear in simple statistics,
and so carry non-trivial information. 

\section{The initial (or pre-hadronic) stage} 
\subsection{Two main scenarios}
Although  we have now reached
some understanding of the {\bf hadronic stage} of the
evolution,
from chemical to kinetic freeze-out,  we 
know very little about
{\it   what happens before it}.
Ignoring  details\footnote{
E.g.  disregarding purely hadronic models, because particle
composition
  cannot be explained by reactions
in hadronic phase. }
, let me
emphasize 
 two major points of view
present on the market: (i) {\bf The QGP scenario}
(hopefully our future Standard Model), based on a picture of quick
matter equilibration; and (ii) {\bf string scenario}, 
represented by most traditional Lund-type event generators like Venus
and RQMD,
or reggion-based approaches like DPM. 
A positive feature of the former approach is  obviously its
conceptual simplicity, while the latter
has direct connection to pp and
pA phenomenology. $Both$ can explain why the initial EOS is  soft enough not to contribute
much to the radial flow.

{\bf Particle composition} would be a stronghold of thermodynamics, if
not for the fact that in pp and $e^+e^-$ collisions
it is also possible to use successfully
statistical models. The  difference 
 is however seen in the
 {\bf strangeness} production, much suppressed in
pp and $e^+e^-$ but not
 in AA\footnote{
Some studies \cite{Raf}
even find that strangeness is slightly enhanced compared to the
equilibrium value, partly due to Coulomb effects.
}.  The QGP scenario explains it naturally, while the
string one needs ``color ropes'' (or other exotic devices) to explain
the strangeness.

 Significant experimental efforts have been made to locate the transition
 between two extremes, the pp-type and heavy-ion-type regimes of
 strangeness production. Excellent data from WA97,NA49 
for $\Lambda,\Xi,\Omega$ and their anti-particles found that
strangeness has
{\it little dependence on centrality}. Lighter ion data obtained before also
confirm the impression that  strangeness production
is {\it basically constant} throughout the  SPS domain,
with the transition being somewhere in the AGS domain.  
To me it indicates that strangeness 
``un-suppression'' is clearly related with 
 the approach of the hadronic phase boundary,  
production of the mixed phase with at least $some$ QGP. 

We will separately discuss
 $\rho$ melting  and $J/\psi$ suppression below: but let me 
now emphasize their
contrasting  A  dependence.
CERES data on dilepton enhancement below $\rho$ show comparable effect in SAu and
PbAu, while the NA50  $J/\psi$ suppression is drastically different. Why is it
so? It is a must to see if the energy dependence of two effects is
different as well.

  By adding $\psi'$
 in between, and decomposing $J/\psi$ into 
$\chi$ and proper $J/\psi$ parts, one would get a whole sequence of
``melting'' phenomena,
happening as matter becomes hotter/denser. We will 
see at RHIC how members of the $\Upsilon$ family do the same later on. 
Which of them is the QGP signal then? Well, this
depends on details which we still have to 
work out.
 
\subsection{Dileptons } 

As ``penetrating
probes'' \cite{Shu_78}, dileptons suppose to tell us 
the story of
 ``melting'' of all vector mesons ($\rho,\omega,\phi,J/\psi,\Upsilon$),
replaced by  radiation from 
thermal quarks.
All SPS dilepton experiments (HELIOS-3,CERES,NA50)
 see significant
dilepton enhancement (compared to ``trivial sources''),
being stronger at small  $p_t$,  indicating
 matter effects. 

The first important issue is the origin of the
 enhancement observed by NA50 at $M_{\mu^+\mu^-}\sim 2 GeV$. 
It was suggested that it is due to {\it
enhanced charm production},
but another (and, in my mind, much more probable) explanation is
the {\it thermal QGP emission} \cite{Shu_78} . The QGP-like rate  reproduces
HELIOS3 data \cite{LiGale}, and preliminary estimates \cite{RappShuryak} show it
works for NA50 data as well.

The origin of a
  qualitative change of the shape of the vector spectral density
  $\rho_v(M)$ for
$M< 1 GeV$ observed by CERES
 was discussed here
 in detail \cite{rapp},  
 let me therefore address only one central question:
 to what extent does the observed
``$\rho$
melting'' indicate an approach to chiral symmetry restoration?

 In-matter $\rho$ width
is significantly increased,
 mostly by re-scattering on nucleons.
The non-trivial fact is: even for small  
masses $M=.3-.6 GeV$ the 
  rate (obtained in a complicated
hadronic calculation \cite{rapp})
 happen to be rather close to  the
 ``partonic'' or QGP rate,  corresponding to
free $\bar q q$ annihilation in the heat bath. It tells us 
that the interaction between  $\bar q$ and  $q$ in the vector
channel is becoming weak. 
  $If$ axial spectral density is  modified similarly as well,
the finite-T Weinberg-like sum rules \cite{Kapusta_Shuryak} demand that the
chirally-odd quantities in its r.h.s. become small, which means
 chiral symmetry restoration.  $If$ the axial spectral density remains
different,  chiral symmetry is still broken.
 Although we cannot
access it directly, one may still look for Dalitz-type decays of $a_1$
 \cite{SH_dileptons}.

\subsection{  $J/\psi$, $\psi'$  suppression } 

 Let me start with a brief comment on
 the $\psi'$ suppression.
 The NA50
  data show that the $\psi'/\psi$ ratio stop falling and is
  stabilized ata  small value, about 4\%.
 An explanation suggested in \cite{SSZ}   is:  all
 $\psi'$ are killed first, but then  are re-created from $J/\psi$.

 The central NA50 finding  is of course a
 statement that $J/\psi$ suppression for central 
 (b$<$ 8 fm) PbPb  collisions
 is different from  extrapolations based on
pA and SA collisions. 
Several mechanisms of this suppression were proposed:
 (i)  gluonic photo-effect \cite{Shu_78,Peshkin_KSN};
 (ii) these states  simply do not exist in QGP 
due to Debye screening  \cite{MS};
 (iii) hadronic co-movers kill them \cite{comovers}; 
 (iv) non-monotonous variation of the QGP lifetime, due to the ``softest point'' \cite{ST_psi}.

New data reported at this meeting have clarified the situation
for the most central collisions: using now a very thin target,
it was found that the 1996 data suffered from multiple interactions.
In fact there is a significantly stronger suppression at small $b$,
making the two component picture with separate $\chi$ and $\psi$
thresholds more probable\footnote{And the idea (iv), 
according to which suppression
may even become weaker at smaller b less likely.}.
It is desirable to make another step, increasing the
 density and/or the famous variable L:
only deformed U provides an opportunity here.

  What else can be done to discriminate
 experimentally these ideas? In particularly, how can we
tell whether suppression happened quickly or took a longer time?
  The old idea is to study suppression dependence on $p_t$.
Unfortunately changing $p_t$ we  also change the kinematics:
e.g. destruction by
   gluons or hadrons goes better if  $p_t$ grows.

  Maybe a better idea \cite{HeiMat} is to 
  use the azimuthal dependence of the suppression.
 Instantaneous suppression should show $no$ asymmetry,
but if it takes a few fm/c
  the anisotropy should show up. The problem is the
initial ``almond'' at $b<$ 8 fm is not very anisotropic,
and for larger $b$ there is no anomalous suppression.
(Here too the deformed U can help.)

\subsection{$\phi$-related puzzles}
 $\phi$ is a little brother of  $J/\psi$, but its production in AA
is $enhanced$ rather than suppressed, as compared to NN. Whatever
effects are killing $J/\psi$'s, $\phi$ is re-created
 because strangeness is
close to equilibrium at chemical freeze-out.

  The first puzzle is the {\it apparent absence of $\phi$ in-matter modification.} As argued
\cite{phi}, 
 even  modest modification of kaons should strongly
(by factor 2 or so)
increase the width of  $\phi$  decay
inside the fireball. Non-negligible fraction of $\phi$, up to a half,
 should decay in-matter, 
while experimentally (see e.g. excellent NA49 data presented here by C.Hohne)
  $no$ $\phi$ modification is seen at all! 

  The second puzzle (already mentioned in the flow section)  is that  
 the $m_t$  slopes of $\phi$ spectra measured by NA49 in KK channel
 are large,  close
    to those of the $p,\bar p$ (particles of similar mass). It
    suggests
that somehow   $\phi$  participates
    fully in the radial flow\footnote{Again, ellipticity may give a clue here.}.
 How is it possible, with its small
 re-scattering cross section?
  New NA50 data on $\phi$ reported at this conference\cite{Gerschel}
   have a different  slope than NA49, only about 220 MeV.
 If extrapolated from larger $p_t$ (where NA50 data are) to small
ones, they go well above the NA49 ones. 

 All puzzles may  be explained if absorption destroy 
 K from most in-matter decays\footnote{Or their $refraction$
 in a collective potential.}, more so at low $p_t$. 
Obviously, there should  be no  missing  $\phi$ 
 in dilepton channel, and so such experiments should see the $\phi$  missing
from KK channel at low $p_t$. (Phenix at RHIC will have good
resolution in both channels, so it should eventually clarify the issue.) 

In summary: as a first (indirect) sign
of  missing (modified?) $\phi$ we have evidences for the unusual
change of its $m_t$ slope. If able to cover smaller $p_t$, 
    NA50 should see 
  $\phi$ spectra which are  {\it different} from NA49,  including
those which decay inside the fireball. 

\section{Conclusions and suggestions}

\subsection{Little Bang versus the Big one}


 It is always fun to notice parallels to
cosmology. There are many  methodic similarities, as well
as  amusingly
close timing of some recent developments.

 The first obvious connection between
the ``Little Bangs" in AA collisions
and the ``Big Bang" is that both
are violent explosions. Expansion of the created
hadronic fireball  approximately follows the Hubble law,
$v\sim r$, although
anisotropic one.
 The $final$  velocities, the Hubble constant and $v_t$, have
been
a matter of debates few years ago, but now are believed to be
fixed (at say 10\% 
level).
The next important issue 
  in both cases is the {\it acceleration history},
 needed to shed some light on  the fundamental EOS.
Cosmologists use distant supernovae to access flow long ago: 
we use $\Omega^-$ to learn what was the flow at ``mid-time''
 (5-10 fm/c).

The last point: angular anisotropy of flow and
its fluctuations.  Amazingly accurate 
 measurements of the microwave background have found a dipole
anisotropy (due to our motion relative to ``ether'')
  and
tiny ($\sim 10^{-5}$) fluctuations with  angular
momentum
$l\sim 100$ due to frozen plasma oscillations, from  the freeze-out
stage in which the QED plasma was
neutralized into ordinary atomic matter. 
In the Little Bang
we have $average$
 dipole and elliptic flows of a few percent. 
 $Fluctuating$
 higher harmonics are not analyzed yet:  there must  be some trace of
  ``frozen QGP oscillations''  as
 well. True, cosmologists have much more photons, but they
 are restricted to {\it only one event}, while
 we have millions of them!

\subsection{Using deformed nuclei (U):  An old idea with a new twist }

An old idea\footnote{Kind of a folklore of our field, the only written
version I found is a memo
  written by P.Braun-Munzinger  to BNL.} is to
select head-on (long-long) collisions, by triggering on maximal
number of participants $N_p$.
Because of larger A and $deformation$, the gain in energy density   
can realistically reach 35-40\% \cite{Shu_U}, which is
 important e.g. for the
 $J/\psi$ suppression studies.

The main finding of my recent
studies of  UU collisions \cite{Shu_U} is however a possible virtue
of ``parallel" collisions, in which
both long axis are orthogonal to the beam. Using 
  $two$ control parameters, the number of participants and 
ellipticity\footnote{The measured  deformation of spectra, $v_2$, is proportional
to this initial deformation (with EOS-depending coefficient!) 
 and should have similar distribution. }, one can effectively separate those.
As one can see from Fig.\ref{fig_elliptic}(b),
unlike  PbPb collisions, the UU ones provide a range 
of deformations, with long-to-short ratio reaching about 1.3.
(Those correspond to collisions with two long directions 
parallel to each other and orthogonal to the beam).
It is comparable to the deformation reached for mid-central
collisions of spherical
nuclei, but now for much larger and denser system.

  As mentioned several times in this talk, it may
help to 
clarify many issues, such as presence of
the QGP push   in elliptic flow at SPS, a time scale
 of the  $J/\psi$ suppression, etc. One more example are
corrections to hard processes, like ``shadowing" due to
initial state re-scattering or  ``jet quenching" due to
final state.
Selecting two geometries, head-to-head and ``parallel",
one can change the longitudinal to transverse size ratio from 1.3 to
1/1.3, a significant level arm to tell the difference.

\subsection{ Experimental goals}

It is  clear that there can only be a
finite time-span for the SPS heavy ion program, so we have to be
``picky". I think the following list includes only experiments
which are a complete ``must''.

\begin{itemize}
\item{--} It is not likely this energy region would be
studied later, so we better be sure no qualitative phenomenon  is
missed.
For  years I advocated measuring  SPS excitation function 
looking for the ``softest
point", and  we will have the 40 GeV run soon.
Now  excitation
 function of $v_2$ became an important issue, with a potential
 to see ``the QGP push'' at SPS.
 Another compelling argument for a scan is
 hunting for the tricritical point of QCD: 
finding it would be a major breakthrough, going to textbooks etc.
\item{--}
 The nature of the dilepton excess for $M\sim2 GeV$ found by NA50
 should be understood. If it is indeed
charm enhancement, up to factor 3 for central PbPb, then the $J/\psi$ 
suppression issue
is much more serious.  If it is QGP radiation, it should be studied more.
 The number one hadronic measurements which 
remains to  be done
is therefore {\it direct observation of charm} by D's.
\item{--} $J/\psi$ $suppression$ is high priority, as
the only observable in which relatively sharp centrality dependence.
 By changing the
beam (A,collision energy) one should test
whether the variation seen is or is not related to 
 fixed energy density. Time-scale of the suppression
can be accessed by studying
its  $p_t$ dependence or
``ellipticity''.
\item{--}
CERES: significant improvement of signal/background ratio is expected
from its recent upgrade, leading to clear separation
of the $\omega$
from the $\rho$ peak, as well as independent look at the $\phi$ shape. 
If it works out as expected, new CERES would be an excellent
tool to study  dramatic 
 in-matter modification (rather than just enhancement or suppression)
of vector resonances. It is also significantly
 statistics-limited experiment, 
 deserving running time
as much as possible.
\end{itemize}


\begin{thebibliography}{20}
\bibitem{RSSV2} R. Rapp, T. Sch\"afer, E. V. Shuryak and M. Velkovsky, Submitted to Ann.Phys.  hep-ph/9904353. 
\bibitem{RS} See talks by K.Rajagopal and T.Schaefer, this volume.
\bibitem{super}D.~Bailin and A.~Love, Phys. Rep. 107,  325 (1984)
\bibitem{super_inst} R. Rapp, T. Sch\"afer, E. V. Shuryak and M. Velkovsky,
hep-ph/9711396; Phys.Rev.Lett., 81:53-56,1998;
M. Alford, K. Rajagopal and F. Wilczek, hep-ph/9711395,
Phys. Lett. {\bf B422} 247 (1998). 
\bibitem{ARW2} M.~Alford, K.~Rajagopal and F.~Wilczek, hep-ph/9804403,
also talks by K.~Rajagopal, this volume.
\bibitem{Son} D.T. Son, Phys.Rev.D59:094019,1999;hep-ph/9812287
\bibitem{SRS} M.Stephanov, K.Rajagopal,
E. Shuryak.Phys.Rev.Lett.81:4816-4819,1998; hep-ph/9806219 and 9903292,
see also talks by K.Rajagopal and
M.Stephanov this volume.
\bibitem{CR}J.Cleymans, K. Redlich, nucl-th/9903063, and  talk by
J.Cleymans, this volume.
\bibitem{UrQMD} L.V. Bravina et al J.Phys.G25:351-361,1999, 
 nucl-th/9810036, and the talk by L.Bravina, this volume.
\bibitem{HW} See talk by U.Wiedemann, this volume.
\bibitem{ES_72} E.Shuryak,
  Phys.Lett.B42 (1972) 357
\bibitem{Cv}
L. Stodolsky,  Phys. Rev. Lett. {\bf 75} (1995)
  1044. E. V. Shuryak, Phys. Lett. {\bf B423} (1998) 9.
\bibitem{BH} G.Baym and H.Heiselberg,  nucl-th/9905022 
\bibitem{Mrow} S. Mrowczynski, Phys.Lett.B (in press),  nucl-th/9901078
\bibitem{SZ}E.Shuryak and O.V.Zhirov, Phys.Lett.89B (1979) 253
\bibitem{HS}C. M. Hung and E.Shuryak,
     Phys.Rev.Lett. 75 (1995) 4003,
 Phys.Rev.C57:1891-1906,1998 and
hep-ph/9709264; E.Shuryak, Proceedings of QM97, Nucl.Phys.A638 (1998) 207.
 \bibitem{deuteron} 
 J.L. Nagle, B.S. Kumar, D. Kusnezov, H. Sorge, R. Mattiello,
 Phys.Rev.C53:367-376,1996
 \bibitem{Sorge_omega} H. van Hecke, H. Sorge, N. Xu.  Phys.Rev.Lett.81:5764-5767,1998 
 nucl-th/9804035  
\bibitem{resonancegas} E.Shuryak, Sov.J. of Nucl. Phys.
16 (1972) 395;
R. Venugopalan and M. Prakash, Nucl. Phys. {\bf A546}
(1992) 718.
\bibitem{Ollitrault}J.Ollitrault
  Phys. Rev. \bf D46\rm, 229(1992); Phys. Rev. \bf D48\rm,
  1132(1993), and talk at QM99.
\bibitem{Sorge} H. Sorge, Phys. Rev. Lett. \bf 78 \rm
  2309(1997) and   \bf 82 \rm 2048 (1999) 
\bibitem{Danielewicz}P. Danielewicz et al
 Phys.Rev.Lett.81:2438-2441,1998: nucl-th/9803047 
and talk by P.Danielewicz this volume.
\bibitem{vanHove} L. Van Hove (CERN).Z.Phys.C21:93,1983
\bibitem{NA49} see talk by A.Poskanzer, this volume.
\bibitem{Stachel} J.Stachel, (INPC 98, Paris), nucl-ex/9903007 
\bibitem{TS} D.Teaney and
    E.Shuryak, nucl-th/9904006.See also
 RHIC predictions, this volume.
\bibitem{Raf} J.Letessier and J.Rafelski,
  hep-ph/9807346
\bibitem{Shu_78}E.Shuryak, Phys.Lett.78B:150,1978 Sov.\ J.\ Nucl.\ Phys. 28,\ 408\ (1978).
\bibitem{rapp}R. Rapp, G. Chanfray, J. Wambach,
 Nucl.Phys.A617:472-495,1997: hep-ph/9702210;
 talk by R.Rapp this volume.
\bibitem{RappShuryak}R. Rapp and E.V.Shuryak, NA50 dileptons as QGP radiation, in progress.
\bibitem{Kapusta_Shuryak}J.I. Kapusta and E.Shuryak,  Phys.Rev.D49:4694-4704,1994 
\bibitem{SH_dileptons} C.M. Hung, E.V. Shuryak, Phys.Rev.C56:453-467,1997: hep-ph/9608299
\bibitem{LiGale}G.Q. Li, C. Gale; Phys.Rev.Lett.81:1572-1575,1998: nucl-th/9805052
\bibitem{phi} D.Lissauer and E.Shuryak, Phys.Lett.B 253 (1991)
15. M.Asakawa and C.M.Ko, Phys.Lett.B322 (1994) 33,
D.Seiberg and C.Gale, Phys.Rev.C52 (1995) R490.
\bibitem{Gerschel}see talk of N.Willis, this volume.
\bibitem{SSZ} H.Sorge, E.Shuryak, I.Zahed 
 Phys.Rev.Lett.79:2775-2778,1997, hep-ph/9705329
\bibitem{Peshkin_KSN}
M.Peskin,Nucl.Phys. B156 (1979) 365;
  D. Kharzeev and H. Satz,  Nucl.\ Phys.\ A590,\ 515c,(1995)
\bibitem{MS} T. Matsui and H. Satz, Phys.\ Lett.\ 178B,\ 416\ (1986).
\bibitem{comovers} S.Gavin and R.Vogt,
Phys.Rev.Lett.78:1006-1009,1997; D.E. Kahana and S.H. Kahana,
nucl-th/9808025. Armesto, A. Capella, E.G. Ferreiro
 Phys.Rev.C59:395-404,1999: hep-ph/9807258 
\bibitem{ST_psi}E. Shuryak and D. Teaney, Phys.Lett.B430:37-42,1998 
\bibitem{HeiMat} H.Heiselberg and R.Mattiello: nucl-th/990100
\bibitem{Shu_U} E. Shuryak,High energy collisions of strongly deformed nuclei:
 An old idea with few new twists, in progress.
\end{thebibliography}
\end{document}